\documentclass[jap,twocolumn,nopacs,amsmath,letter]{revtex4}
\usepackage{graphicx}

\newcommand{\Journal}[4]{#1 \textbf{#2}, #3 (#4)}

\begin{document}

\title{Current-Driven Switching in Magnetic Multilayer Nanopillars}

\author{S. Urazhdin}
\author{Norman O. Birge}
\author{W. P. Pratt Jr.}
\author{J. Bass}
\affiliation{Department of Physics and Astronomy, Center for
Fundamental Materials Research and Center for Sensor Materials,
Michigan State University, East Lansing, MI 48824}


\begin{abstract}
We summarize our recent findings on how the current-driven
magnetization switching in nanofabricated magnetic multilayers is
affected by an applied magnetic field, changes of temperature,
magnetic coupling between the ferromagnetic layers, variations in
the multilayer structure, and the relative rotation of the layers'
magnetizations. We show how these results can be interpreted with
a model describing current-driven excitations as an effective
current-dependent magnetic temperature.

\end{abstract}
\maketitle

\section{Introduction}

An exchange-based mechanism for current-driven switching of
magnetization was predicted by Slonczweski~\cite{slonczewski} and
Berger,~\cite{berger} and later observed
experimentally.~\cite{tsoiprl,cornellscience} Recent research has
been directed towards better understanding of the current-driven
excitation mechanism, adequate description of the magnetic
dynamics, and optimization of magnetic devices e.g. to decrease
the switching current $I_s$ for possible applications in magnetic
memory.
~\cite{slonczewski2,berger2,waintal,sun,tsoinature,cornellorig,cornellapl,grollier,cornelltemp,cornellquant,
zhang,wegrowe,tsoiprl2,
stiles,kovalev,sun2,chien,kent,shpiro,myapl,myprl,mancoff,zhang2,
cornellmicrowaves,kochsun,myjmmm,iswvsmr,mytheory}

We used the giant magnetoresistance effect (MR) to study
current-driven switching in nanofabricated magnetic F$_1$/N/F$_2$
trilayers (nanopillars). We studied how the current-driven
switching is affected by variations of the magnetic field $H$,
ambient temperature $T_{ph}$,~\cite{myprl} coupling between
magnetic layers,~\cite{myapl,myjmmm} electron spin-flipping in the
spacer N or outside the trilayer, and mutual rotation of the two
layers' magnetizations~\cite{iswvsmr}. We interpret our
experimental results for currents that are not too large in terms
of the recently proposed effective temperature
model.~\cite{mytheory}

Our samples were made with a multistep process described
elsewhere.~\cite{myapl} Below, all thicknesses are in nm. The
basic samples had structure
Cu(80)/F$_1$(20-30)/N(10-15)/F$_2$(2-6)/Cu(2-5)/Au(150). F was
either Co or Py=Ni$_{84}$Fe$_{16}$. Although Co is commonly used
in studies of current-driven switching, Py has the advantage of
small crystalline anisotropy and magnetostriction, allowing
reproducible measurements at both 295~K and 4.2~K. The bottom
Cu(80) layer was the extended lead, N, F$_2$ and Cu(2-5) were
patterned into an elongated shape with dimensions $\approx
(130-140)\times (60-70)$~nm$^2$, and Au(150) was the top lead. In
all the samples, except for the AF-coupled ones (see
Section.~\ref{magncoupl}), F$_1$ was left extended to minimize the
effect of dipolar coupling on the current-driven
switching.~\cite{cornellapl} Details of specific samples will be
given with their data. We measured $dV/dI$ with four-probes and
lock-in detection, adding an ac current of amplitude 20--40~$\mu$A
at 8~kHz to the dc current $I$. Positive current flows from F$_1$
to F$_2$. $H$ is in the film plane and (except for the angular
dependence studies) along the nanopillar easy axis.

\section{Switching with varied H and $T_{ph}$}

\begin{figure}
\includegraphics[scale=0.4]{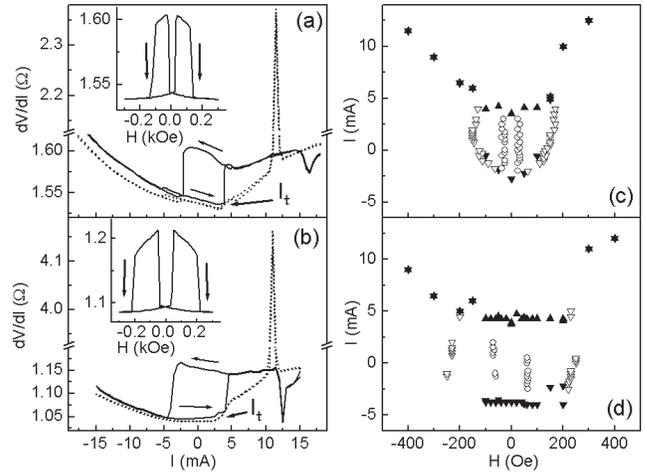}
\caption{\label{fig1} (a) Switching with current at 295~K in an
uncoupled Py/Cu/Py trilayer. Solid line: $H=50$~Oe, dashed line:
$H=-500$~Oe. Arrows mark the scan direction. $I_t$ is the
threshold current as defined in the text. Inset: MR dependence on
$H$ at $I=0$. (b)  Same as (a), at 4.2 K. (c),(d) Magnetization
switching diagram, extracted from the current-switching at fixed
$H$ (solid symbols), and field-switching at fixed $I$ (open
symbols): (c) at 295 K, (d) at 4.2 K. Downward triangles: AP to P
switching, upward: P to AP switching, circles: switching of the
extended Py(20) layer. The reversible switching peaks are marked
by coinciding upward and downward triangles. From Urazhdin {\it et
al.}~\cite{myprl}}
\end{figure}

Figs.~\ref{fig1}(a,b) and insets show the variations of $dV/dI$
with $I$ and $H$ for a Py(20)/Cu(10)/Py(6) sample with uncoupled
Py layers at 295~K (1a) and 4.2~K (1b), for $H=50$~Oe (solid
curves) and $H=500$~Oe (dashed curves). The dependencies are
similar to those seen in Co/Cu/Co samples.~\cite{cornellapl,myapl}
At small $H$, the magnetization switches hysteretically to a
higher resistance AP state at a large enough positive current
$I_s^{P\to AP}\equiv I_s>0$, and to a low resistance P state at
$I_s^{AP\to P}<0$. At larger $H$, the switching step turns into a,
often much higher, nonhysteretic peak. The asymmetric
$I$-dependence is a signature of the effect of the current on the
magnetization, different from the Oersted field.
Figs.~\ref{fig1}(c,d) show the switching diagrams at 295 K and 4.2
K, extracted from data such as those in Figs.~\ref{fig1}(a),(b),
obtained both by varying $I$ at fixed $H$ and $H$ at fixed $I$. As
expected, both the reduced magnetization and thermal activation
result in smaller switching currents and fields $H_s(I=0)$ at
295~K. The only other major difference between the 295~K and 4.2~K
data in Figs.~\ref{fig1}(a,c) is the rounding of the 295~K
hysteretic region at $I<0$. In contrast, the hysteretic region at
4.2~K (Fig.~\ref{fig1}(d)) is almost square.

\begin{figure}
\includegraphics[scale=0.4]{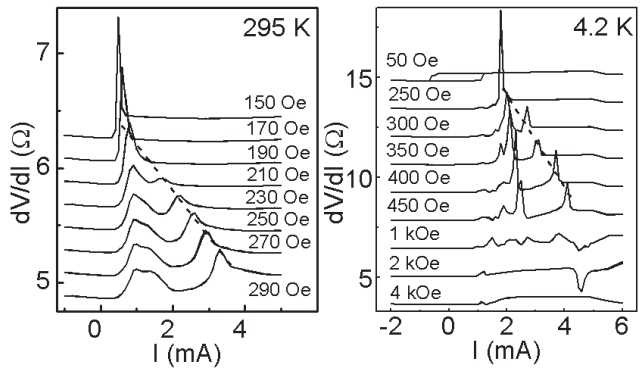}
\caption{\label{fig2}  Differential resistance of a
Py(20)/Cu(10)/Py(3.5) uncoupled trilayer at the marked values of
$H$, at 295~K (left) and 4.2~K (right). Curves are offset for
clarity. Dashed lines follow the reversible switching peak.}
\end{figure}

In the reversible switching regime, both the 295~K and 4.2~K
current-dependencies in Fig.~\ref{fig1} show a nearly linear rise
of $dV/dI$ above a threshold current $I_t$ (labeled in
Figs.~\ref{fig1}(a,b)). In some samples, this rise displays
structures, clearly distinguishable from the switching peak due to
their very weak dependence on $H$. Fig.~\ref{fig2} shows data for
a small Py(20)/Cu(10)/Py(3.5) sample, with dimensions estimated at
$50\times100$~nm$^2$. At 4.2~K, the linear rise is resolved into
several peaks. They appear only to the left of the reversible
switching peak, as the latter moves to higher $I$ with increasing
$H$ (shown by a dashed line). At 295~K, the peaks are replaced by
a smeared non-monotonic structure in $dV/dI$. Based on the data
shown in Figs.~\ref{fig1} and \ref{fig2}, and 9 other similar
Py/Cu/Py samples studied,~\cite{myprl} we summarize the aspects of
the threshold behavior: i) $I_t$ is close to the small-H value of
$I_s^{P\to AP}$ at 4.2~K.  The correspondence $I_s\approx I_t$ is
not universal, e.g. in magnetically coupled samples the switching
may occur at $I<0$ (see Section~\ref{magncoupl}). ii) Both $I_t$
and the sometimes observed structures in $dV/dI$ at $I>I_t$ are
only weakly dependent on $H$, as compared to the reversible
switching peak. The 4.2~K data in Fig.~\ref{fig2} for $H=4~kOe$
give $I_t\approx 1.0$~mA, only 25\% larger than $I_t\approx
0.8$~mA at $250$~Oe. iii) The height and inverse width of the
reversible switching peak in $dV/dI$ are correlated with the
structure in $dV/dI$ at $I>I_t$: for example, in Fig.~\ref{fig2}
(the 295~K data), the reversible peak is tall and narrow when it
is on the rising slope of the 'hump', but becomes wide and nearly
disappears when on the trailing slope.

We discuss first the nature of the threshold behavior starting at
$I_t$, and then the reversible switching peak at higher $I$. $I_t$
has been identified as the onset of large amplitude excitation of
the patterned Py layer, as the magnetic energy provided by the
current exceeds the linear magnetic damping rate.~\cite{myprl}
Starting at $I_t$, there is a strong increase of excitation
amplitude, and the highly excited magnetic state is manifested by
increases in $dV/dI$, as illustrated in
Figs.~\ref{fig1},~\ref{fig2}. In all of our Co/Cu/Co samples, and
most of our Py/Cu/Py samples, the excitations result in a nearly
linear increase in dV/dI above It (see Fig.~\ref{fig1}, and high H
curves in Fig.~\ref{fig4}(b)).  In contrast, in Py/Cu/Py samples
with small lateral dimensions, at 4.2~K the excitations appear as
a series of peaks in $dV/dI$ above $I_t$ (Fig.~\ref{fig2}),
corresponding to step increases in resistance $V/I$, with details
that vary from sample to sample.  We attribute these peaks to
irregularities in sample shapes, resulting in a complicated
dependence on $I$ of both the magnetic damping rates and the
distribution of excitations among different magnetic modes.  In
larger samples, the shape irregularities are less significant,
giving a smoother $dV/dI$.

A different, quantum-mechanical origin has been proposed for the
peaks, interpreted as the excitation threshold in point-contacts
on magnetic multilayers.~\cite{tsoiprl} This threshold would be
due to the requirement for matching of the current-driven
spin-accumulation with the lowest energies of the magnetic
excitations, $\Delta\mu=\hbar\omega$. In nanopillars, the quantum
threshold value would double when $H$ is increased from $0$ to
$\approx 1$~kOe. Such a rapid increase is inconsistent with our
data. In addition, even at 4.2~K, thermal smearing with
$kT_{ph}>>\hbar\omega\approx 10\ \mu$eV would completely smear out
the quantum threshold. Thus, experiments rule out the quantum
nature of the excitation threshold in nanopillars. We note that at
large $H>2\pi M\approx 5$~kOe, both the classical and
quantum-mechanical thresholds depend linearly on $H$, and
$\hbar\omega$ may become comparable to $kT$. In this limit, the
quantum threshold may become important.

\begin{figure}
\includegraphics[scale=0.4]{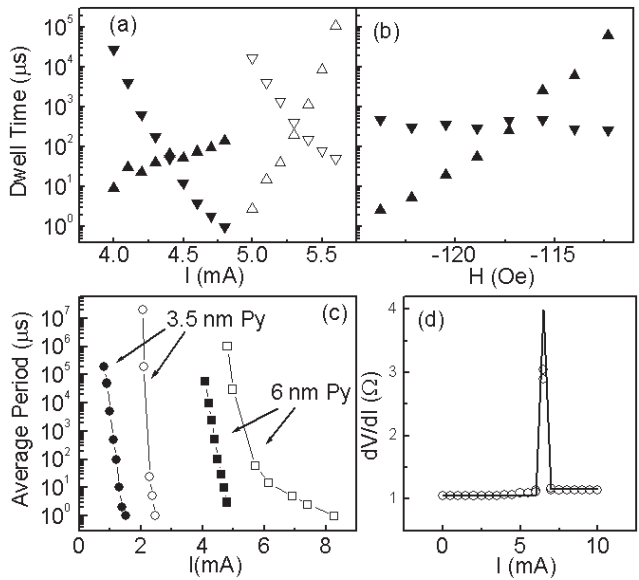}
\caption{\label{fig3} (a) The dependence of the P (downward
triangles) and AP (upward triangles) dwell times on $I$, for a
Py(20)/Cu(10)/Py(6) uncoupled sample. Open symbols:
$T_{ph}=4.2$~K, $H=-335$~Oe, filled symbols: $T_{ph}=295$~K,
$H=-120$~Oe. (b) Dependence of P (downward triangles) and AP
(upward triangles) dwell times on $H$ at $I=4.4$~mA, $T=295$~K.
(c) Current dependence of the average telegraph noise period. $H$
was adjusted approximately linearly with $I$, so that the average
dwell times in AP and P states were equal. Solid circles:
Py(20)/Cu(10)/Py(3.5) at 295~K, $H=$-93 to -121~Oe, open circles:
same sample at 4.2~K, $H=$-300 to -315~Oe, solid squares:
Py(20)/Cu(10)/Py(6) at 295~K, $H=$ -113 to -123~Oe, open squares:
same sample at 4.2~K, $H=$-300 to -450~Oe. (d) Circles: $dV/dI$
{\it vs.} $I$ at $H=-0.3$~kOe, 4.2~K. Solid curve: a calculation,
as described in the text, with $I_0=6.45$~mA, and
$\alpha+\beta=27$~mA$^{-1}$ obtained from (a).}
\end{figure}

The reversible switching peak in $dV/dI$ is different from the
threshold $I_t$ and the peaks we associate with magnetic
excitations. Time resolved measurements showed that it is a
consequence of telegraph noise with random distribution of dwell
times in the P and AP states.~\cite{myprl,myapl}
Fig.~\ref{fig3}(a) shows the variations of average dwell times
$\tau_P(\tau_{AP}$) in the P(AP) state with $I$ for a
Py(20)/Cu(10)/Py(6) sample at 295~K and 4.2~K. $\tau_P$ decreases
as $I$ increases, but $\tau_{AP}$ increases. These variations have
a similar form at 295~K and 4.2~K, and are also similar to the
variations in Co/Cu/Co.~\cite{cornelltemp} Fig.~\ref{fig3}(b)
shows that at a fixed $I$, $\tau_P$ increases and $\tau_{AP}$
decreases as $|H|$ is increased (shown for 295~K).
Fig.~\ref{fig3}(c) shows that, when both $I$ and $H$ are increased
so as to hold $\tau_P=\tau_{AP}$, the average period of the
telegraph noise decreases exponentially with similar slopes at
295~K and 4.2~K, down to the 1~MHz bandwidth limit of our setup.

We now show how a peak in $dV/dI$ at $\tau_{P}\approx\tau_{AP}$
can be derived from the variations of $\tau_P$, $\tau_{AP}$ with
$I$. For a fixed $H$, the average voltage across the sample is
\begin{equation}\label{voltage}
V(I)=I\left[\frac{R_{AP}\tau_{AP}+R_{P}\tau_{P}}{\tau_{P}+\tau_{AP}}\right],
\end{equation}
with $\tau_{P}(I)\approx\tau_0\exp[-\alpha(I-I_0)]$,
$\tau_{AP}(I)\approx\tau_0\exp[\beta(I-I_0)]$, as in
Fig.~\ref{fig3}(a). We define $I_0$, $\tau_0$ by
$\tau_{AP}(I_0)=\tau_{P}(I_0)=\tau_0$. From Eq.~(\ref{voltage})
\begin{eqnarray}\label{dvdi}
dV/dI\approx
\nonumber R_P+\frac{R_{AP}-R_P}{1+\exp[-(\alpha+\beta)(I-I_0)]}+\\
I(R_{AP}-R_P)\frac{\exp[(\alpha+\beta)(I-I_0)](\alpha+\beta)}{(\exp[(\alpha+\beta)(I-I_0)]+1)^2}.
\end{eqnarray}
The first two terms on the right are the resistance $V/I$, giving
a step for the reversible transition from P to AP. For large
$I_0(\alpha+\beta)$, the last term has a maximum value
$I_0(R_{AP}-R_{R})(\alpha+\beta)/4$ at $I\approx I_0$. This term
gives rise to a peak in $dV/dI$ at $I=I_0$ that can be much higher
than $R_{AP}$. Fig.~\ref{fig3}(d) shows a calculation (solid line)
based on the data of Fig.~\ref{fig3}(a), and Eq.~(\ref{dvdi}), for
$I_0=6.45$~mA, and $\alpha+\beta=27$~mA$^{-1}$ extracted from
fig.~\ref{fig3}(a). The calculation is consistent with the $dV/dI$
measurement (circles). We conclude that the reversible switching
peak positions characterize the points $(H,I)$ where
$\tau_P=\tau_{AP}$. The slope of the reversible switching line is
a measure of the telegraph noise variation with $I,H$. From
Eq.~\ref{dvdi}, the height and inverse width of the reversible
switching peak are given by $\alpha+\beta$, which is usually
dominated by $\alpha$, i.e. the dependence of the magnetic
excitation rate in the P state on $I$. This is consistent with the
correlation in Fig.~\ref{fig2}(a) between the reversible switching
peak and the 'bump' associated with magnetic excitations: On the
positive slope of the 'bump' ($H=150,170$~Oe), the excitation
level grows faster with $I$ than on its the trailing slope.
Consequently, $\alpha$ is larger on the rising slope of the bump,
giving a tall and narrow reversible switching peak. The reversible
peak nearly disappears on the trailing slope at $H=190,210$~Oe,
where $\alpha$ is small.

In contrast to $I_t$, the switching peak does not represent an
onset of a physical process, it merely reflects the
current-dependent telegraph noise statistics. The presence of
telegraph noise near the reversible switching line indicates that
both AP and P states are unstable there. Strictly, the stability
diagrams, Figs.~\ref{fig1}(c,d) should be modified to include this
unstable region. In most uncoupled samples, the P-state becomes
unstable at $I\approx I_t$. The instability of the P-state at
$I>I_t$ is thus indirectly manifested in the rise of $R_P$. The
AP-state is unstable at $I$ both below and above the reversible
switching peak. Since $\tau_{P}$ is exponentially smaller than
$\tau_{AP}$, the measurements of $dV/dI$ at $I$ above the
reversible switching peak give values very near $R_{AP}$.

As $I$ and $H$ are increased, $\tau_P,\tau_{AP}$ quickly approach
the intrinsic nanosecond switching timescale (see
Fig.~\ref{fig3}(c)), where telegraph noise is replaced by some
sort of fast magnetic dynamics. Such a transition does not give
any significant changes in the switching peaks in $dV/dI$ (see
e.g. Fig.~\ref{fig2}). Although the bandwidth limitation of our
setup does not allow us to directly probe this regime,
spectroscopic measurements show continuous spectrum from 0~Hz up
to $\approx 1$~MHz, characteristic of incoherent magnetic
dynamics.~\cite{myunpublished} Microwave measurements show peaks
at GHz frequencies on top of a broad background, evidence for some
level of coherence.~\cite{cornellmicrowaves}

\section{\label{magncoupl} Effects of Magnetic Coupling}

\begin{figure}
\includegraphics[scale=0.36]{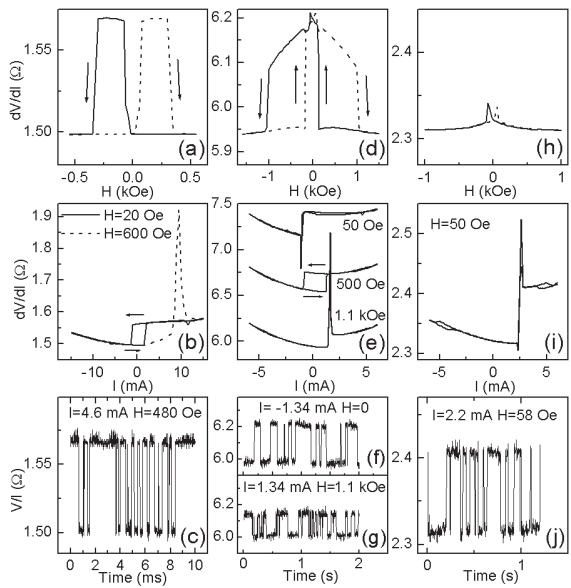}
\caption{\label{fig4} Results for uncoupled (a-c), AF-coupled
(d-g), and F-coupled (h-j) Co/Cu/Co samples at 295~K. In the
hysteretic plots, arrows show the scan direction. (a,d,h) dV/dI
{\it vs.} H at $I=0$. (b,e,i) dV/dI {\it vs.} I at the listed
values of H. In (e) curves are offset for clarity. (c,f,g,j)
Time-resolved measurements of R=V/I at the listed values of H, I.
From Ref.~\cite{myjmmm}}
\end{figure}

Fig.~\ref{fig4} summarizes the differences between uncoupled
(left), AF-coupled (middle), and F-coupled (right)
Co(20)/Cu(d)/Co(2.5) samples at 295~K. For uncoupled samples, we
used d=10, and only patterned the top Co(2.5) and most of the
Cu(10) layer to nanopillar size, leaving the bottom Co(20)-layer
extended. F coupling was achieved by reducing the Cu thickness to
Cu(2.6), near the third RKKY magnetoresistance (MR)
minimum.~\cite{parkin} Dipolar AF coupling was achieved by also
patterning about 10~nm of the Co(20) layer, with $d=6$. Sample
shape variations and interfacial roughness lead to variations in
both AF and F coupling strengths.

Fig.~\ref{fig4}(a)) shows field-driven switching in uncoupled
samples, similar to that in Fig.~\ref{fig1} for uncoupled Py-based
samples. In AF-coupled samples (Fig.~\ref{fig4}(d)), the Co(2.5)
layer is oriented AP to the extended Co(20) layer by the dipolar
field, so only the high resistance AP state is stable at $H=0$. In
contrast to Fig.~\ref{fig4}(a), both the P$\to$AP and AP$\to$P
transitions are now due to switching of the Co(2.5) layer. As $H$
is scanned from a large negative to a large positive value, this
layer switches three times: once to the AP state at negative $H$,
then together with the Co(20) layer at small positive $H$ to stay
in the AP state, and finally to a P state at large positive $H$.
The second transition sometimes produces a weak feature in
$dV/dI$. In contrast, F-coupling causes the magnetizations to
reverse simultaneously at small H, giving only a small feature in
MR at I=0 (Fig.~\ref{fig4}(h)). In H-scans at large enough fixed
$I>0$, F-coupled samples gave 5\% MR, similar to the $I=0$ values
for uncoupled or AF-coupled samples.

Fig.~\ref{fig4}(b,e,i) compares the variations of $dV/dI$ with
$I$. At small $H=20-50$~Oe, applied to fix the magnetization of
the bottom Co layer, the uncoupled sample (Fig.~\ref{fig4}(b),
solid curve) shows hysteretic switching, while the AF-coupled
(Fig.~\ref{fig4}(e), top curve) and F-coupled (Fig.~\ref{fig4}(i))
ones show reversible switching at $I<0$ and $I>0$, respectively.
At larger H, the switching in uncoupled samples
(Fig.~\ref{fig4}(b), dashed curve) becomes reversible, and in
AF-coupled samples (Fig.~\ref{fig4}(e), lower two curves) it first
becomes hysteretic and then reversible again. Time resolved
measurements of resistance at $I,H$ near the reversible switching
peaks show telegraph noise switching between AP and P states
(Fig.~\ref{fig4}(c,f,g,j)). At identical $I$ of opposite signs,
the average telegraph noise periods in the AF-coupled sample are
similar.

\section{Relation Between MR and $I_s$}

\begin{figure}
\includegraphics[scale=0.25]{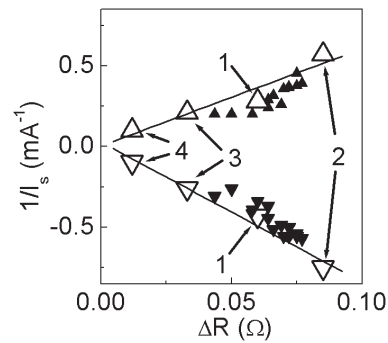}
\caption{\label{fig5} Dependence of $1/I^{P\to AP}_s$ (upward
triangles) and $1/I^{AP\to P}_s$ (downward triangles) on $\Delta
R$ for uncoupled samples. Open symbols: sample types 1 through 4,
as labeled and explained in the text. Solid lines: best linear
fits of data, excluding the angular dependence. The ordinate
intercepts are zero within the uncertainty of the fits.}
\end{figure}

We studied the correlation between MR and $I_s$ in uncoupled
samples with a structure Cu(80)/Py(30)/N(15)/Py(6)/Cu(2)/Au(150).
N=Cu in sample types 1 and 2. In sample types 3 and 4, N was
Cu(9.5)/Cu$_{94}$Pt$_6$(4)/Cu(1.5) and
Cu(5.5)/Cu$_{94}$Pt$_6$(8)/Cu(1.5), respectively. The short 295~K
spin-diffusion length $t_{sf}\approx 6$~nm ($\approx 9$~nm at
4.2~K~\cite{sdlength}) in Cu$_{96}$Pt$_6$ decreases $\Delta
R\equiv R_{AP}-R_P$. We enhanced $\Delta R$ in sample type 2 by
replacing Cu(2) with a Cu(2)/Fe$_{50}$Mn$_{50}$(1)/Cu(2) sandwich.
Fe$_{50}$Mn$_{50}$ is a strong spin-scatterer.~\cite{sdlength} Its
placement between the trilayer and the top lead reduces the
negative effect of the spin accumulation outside the trilayer on
the MR. We also examined the variation of $\Delta R$ and the
switching currents $I_s$ in samples of type 1 while rotating the
magnetization of the Py(30) with a small $H$ in the film plane
(similar results were shown in~\cite{mancoff}).

The results, averaged over at least 7 samples of each type, are
summarized in Fig.~\ref{fig5}, where data for angular dependence
are also shown. Both $I_s^{AP}$ and $I_s^{AP}$ follow an
approximately inversely linear dependence on $\Delta R$,
regardless of the method by which $\Delta R$ was varied.

\section{Effective Temperature Model}

In their pioneering work, Slonczweski~\cite{slonczewski} and
Berger,~\cite{berger} predicted that exchange interaction leads to
current-driven magnetic excitation by spin-polarized current.
Berger considered generation of magnons by spin-flipping of
electrons, driven by the spin accumulation. Slonczewski considered
the electron spin component transverse to the magnetization, which
is absorbed by the ferromagnet due to a combination of
spin-dependent reflection at the interfaces and averaging of the
spin precession phases in the ferromagnet. The resulting torque
drives the magnetic dynamics. This model postulates conservation
of the total magnetic moment, and thus captures only the coherent
(uniform) magnetic dynamics. It can not consistently treat the
excitation of finite-wavelength spin-waves, whose generation is
generally not associated with transfer of angular momentum
transverse to the magnetization (a similar argument is used to
prove that only uniform precession is excited in the transverse
FMR experiments). Similarly, Berger made the approximation that
only the uniform precession is generated by electron
spin-flipping, leading again to coherent magnetic dynamics.

Since the characteristic magnon energies are significantly lower
than the typical conduction electron energies (see the above
discussion of the lack of quantum threshold behavior), we assume
that a large number of magnetic modes are excited by the current.
The populations $n_i$ of the modes with energies $E_i$ can be then
approximately described by a single parameter, an effective
temperature $T_m$, so that $n_i\approx k_BT_m/E_i$ for the
degenerate modes.~\cite{myprl,myapl,mytheory} This approximation
fails when $T_m$ approaches the Curie temperature of the
ferromagnet. We emphasize that this is just a single-parameter
approximation for a generally much more complex, non-thermalized
excitation distribution. Because of the large magnon populations,
the spontaneous (independent of $n_i$) magnon emission can be
neglected. In a ballistic transport approximation~\cite{mytheory}
\begin{equation}\label{effecttemp}
kT_m\approx \frac{kT_{ph}}{1+2peVB/\gamma},
\end{equation}
where $B$ is a constant characterizing the strength of the
exchange interaction, $\gamma$ is related to the Gilbert damping
parameter in the classical Landau-Lifshitz equation, $V$ is the
voltage across the trilayer, and $p$ is the current polarization,
created at the location of F$_2$ by F$_1$, if F$_2$ is removed.
$p>0$($<0$) in the P(AP) state for Py/Cu/Py or Co/Cu/Co trilayers,
and $p=0$ if F$_1$ is removed. Eq.~\ref{effecttemp} is similar to
the expression obtained by Berger in a diffusive transport
approximation (Eq.~(9) in Ref.~\cite{berger2}).

Eq.~(\ref{effecttemp}) diverges at $V\to -\gamma/(2peB)$. We
identify this divergence with the threshold $I_t$ for the large
amplitude of excitations. Eq.~(\ref{effecttemp}) is not applicable
above the threshold, because the conduction electrons are
scattered with a large spin-flip probability ($\approx 0.5$),
nearly independent of $T_m$. For this regime, we have proposed an
empirical relation~\cite{myprl}
\begin{equation}\label{tm}
T_m=T_{ph}+K(I-I_t)\mbox{ for }I>I_t,
\end{equation}
where $K$ is a constant determined by the magnetic relaxation
rate. In the thermal activation model, the dwell times
$\tau_{P,AP}$ are determined by
\begin{equation}\label{expdec}
\tau_{P,AP}=\frac{1}{\Omega}exp\left[\frac{U_{P,AP}}{kT^{P,AP}_m}\right],
\end{equation}
where $\Omega\approx 10^7s^{-1}$~\cite{koch} is the effective
attempt rate, $U_{P,AP}$ is the potential barrier for switching
from the P or AP state. $T_m$ is approximated by
Eq.~(\ref{effecttemp}) for the low-H hysteretic switching, or
Eq.~(\ref{tm}) for the telegraph noise. The areas of the P(AP)
state stability are determined by $\tau_{P(AP)}>1$~s. $U_{P,AP}$
depend on $I$ only through the variation of the magnetization with
$T_m$.

\begin{figure}
\includegraphics[scale=0.4]{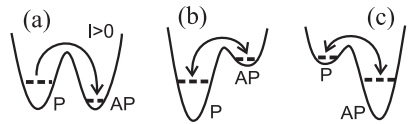}
\caption{\label{fig6} Schematics of current-driven switching, as
explained in the text. Dashed lines indicate $T_m$.}
\end{figure}

Fig.~\ref{fig6}(a) schematically describes hysteretic
current-driven switching in uncoupled samples at small H.  From
Eq.~(\ref{effecttemp}), $T_m^P$ drastically increases when $I$
approaches $I_t$, giving rise to a thermally activated transition
into the AP state at
$kT_m^P\approx\frac{U_P}{ln(t_{exp}\Omega)}\approx
\frac{U_P}{16}$, for the data acquisition time
$t_{exp}=1$~sec/point. In the AP state, Eq.~(\ref{effecttemp})
predicts that the magnetic system weakly cools to $T_m^{AP}<
T_{ph}$, so that at $H<H_s$ it becomes trapped in this state.
Fig.~\ref{fig6}(b) is a schematic for $H>H_s$, giving
$kT_{ph}>\frac{U_{AP}}{16}$. The AP to P switching is now
thermally activated, resulting in telegraph noise at $I>I_s$,
$H>H_s$. Fig.~\ref{fig6}(c) is the schematic for AF-coupled
samples at small $H$. AF coupling increases $U_{AP}$ and reduces
$U_P$, so that P to AP switching is now thermally activated.  At
large enough $I<0$, reverse switching also becomes thermally
activated, giving telegraph noise.  As $H$ grows, the schematic
for AF-coupling becomes 6(a), giving hysteretic switching, and
then 6(b), giving telegraph noise for large enough positive $I$.
Finally, 6(b) is the schematic for F-coupling at all $H$, so that
AP to P switching is always thermally activated and large enough
$I>0$ gives telegraph noise.

The difference between the 295~K and 4.2~K switching diagrams
(Fig.~\ref{fig1}) is qualitatively well described by the effective
temperature model.~\cite{mytheory} The rounding of the 295~K
diagram at $I<0$ is due to enhancement of thermal fluctuations of
magnetization even at small $I$ (see Eq.~\ref{effecttemp}).
Increasing $H$ decreases $U_{AP}$, so that even these weak
excitations can activate the AP$\to$P transition. In contrast, the
almost square 4.2~K diagram reflects the small excitation
amplitude at $I<I_t$, beyond which the excitation rapidly grows,
giving the switching.

We use Eqs.~(\ref{tm}) and (\ref{expdec}) to describe the
dependencies of $\tau_P$ on $I$, $H$, and $T_{ph}$ in
Fig.~\ref{fig3}. Eq.~(\ref{effecttemp}) should be used for
$\tau_{AP}(I,H,T_{ph})$. These equations capture the tendencies
qualitatively well. From the slope of $\tau_P(I)$, the increase
rate of $T_M^{P}$ in Py(20)/Cu(10)/Py(6) samples is estimated at
$\approx400$~K/mA.~\cite{myprl} The thermal activation model
breaks down when $k_BT_m$ becomes similar to $U_P$. This
limitation of the model should be understood in the context of the
actual magnetic dynamics. $T_m\approx U$ in Eq.~\ref{expdec} gives
a switching rate similar to the intrinsic magnetic dynamic rates.
Telegraph noise switching is then replaced with fast fluctuations
of the magnetic moment. The effective temperature concept (defined
for a given magnetic orientation) becomes irrelevant in this
regime.

We note that, although the 295~K and 4.2~K current-dependencies in
Fig.~\ref{fig3}(a) are given for different values of $H$, the
effect of this difference on $\tau_{P}$ should be small, as can be
seen from Fig.~\ref{fig3}(b). Since $U_P$ is large, its relative
variation with $H$ is small, compared to much larger relative
variation of $U_{AP}$. Thus, Fig.~\ref{fig3}(a) can be used to
determine $\tau_P(I)$, while $\tau_{AP}$ is adjusted with $H$.

Finally, to describe the inverse relationship in Fig.~\ref{fig5},
we use the simplest plausible model, in which $\Delta R$ is
proportional to the current polarization $p$ at the location of
F$_2$ due to F$_1$. This crude approach is self-consistent: the MR
is correctly predicted to disappear if F$_1$ is absent. The
current-driven switching occurs at $16k_BT_m\approx U$. From
Eq.~\ref{effecttemp} follows $1/I_s\propto p$, giving the
dependence in Fig.~\ref{fig5}. In this context, the angular
dependence can be understood similarly in terms of the projection
of $p$ onto the direction of the magnetization of F$_2$.

\section{Summary}

We summarize the following important experimental observations for
current effects in magnetic trilayer nanopillars: i) a square
switching diagram at 4.2~K, and rounded at 295~K; ii) an onset
current $I_t$ (closely related to the hysteretic switching current
$I_s$) for a linear rise of $dV/dI$ in larger samples or a series
of peaks in smaller ones; iii) reversible switching, characterized
by telegraph noise with rate both increasing exponentially with
$I$ and shifting with temperature. The reversible switching peak
in $dV/dI$ occurs when the dwell times in the P and AP states are
approximately equal. iv) Reversible switching at small $H$, if
magnetic layers are F- or AF-coupled. v) Inverse linear relation
between MR and $I_s$. We explain the observed behaviors in terms
of thermal activation over a magnetic barrier, with a current
driven effective magnetic temperature.

We acknowledge helpful communications with M.D. Stiles, A.H.
Macdonald, D.C. Ralph, S. Zhang, A. Fert, support from the MSU
CFMR, CSM, the MSU Keck Microfabrication facility, the NSF through
Grants DMR 02-02476, 98-09688, and NSF-EU 00-98803, and Seagate
Technology.

\end{document}